\newif\ifpreprint

\preprintfalse % Enable for double column reprint

\ifpreprint
\documentclass[journal=jctcce,manuscript=article]{achemso}
\else
\documentclass[journal=jctcce,manuscript=article,layout=twocolumn]{achemso}
\fi

\usepackage[T1]{fontenc} % Use modern font encodings

\usepackage{amsmath}
\usepackage{newtxtext,newtxmath}

\usepackage{graphicx}
\usepackage{dcolumn}
\usepackage{braket}
\usepackage{multirow}
\usepackage{threeparttable}
\usepackage{xspace}
\usepackage{verbatim}
\usepackage[version=4]{mhchem} % Formula subscripts using \ce{}
\usepackage{comment}
\usepackage{color,soul}

\usepackage{physics}
\usepackage{mathtools}
\usepackage{xcolor}
\usepackage{xspace}
\usepackage{ifthen}

\usepackage{qcircuit}
\usepackage{bm}
\usepackage{cprotect}

\usepackage{listings}
\definecolor{codegreen}{rgb}{0,0.6,0}
\definecolor{codegray}{rgb}{0.5,0.5,0.5}
\definecolor{codepurple}{rgb}{0.58,0,0.82}
\definecolor{backcolour}{rgb}{0.95,0.95,0.92}

\lstdefinestyle{mystyle}{
    backgroundcolor=\color{backcolour},
    commentstyle=\color{codegreen},
    keywordstyle=\color{magenta},
    numberstyle=\tiny\color{codegray},
    stringstyle=\color{codepurple},
    basicstyle=\ttfamily\footnotesize,
    breakatwhitespace=false,
    breaklines=true,
    captionpos=b,
    keepspaces=true,
    numbers=left,
    numbersep=5pt,
    showspaces=false,
    showstringspaces=false,
    showtabs=false,
    tabsize=2
}

\newcommand*{\qft}{\textsc{QForte}\xspace}
\newcommand*{\pybind}{\textsc{Pybind11}\xspace}

\mathchardef\mhyphen="2D

\newcommand{\cop}[1]{\hat{a}^{\dagger}_{#1}}
\newcommand{\aop}[1]{\hat{a}_{#1}}
\newcommand{\code}[1]{{\normalfont\ttfamily #1}\xspace}

\usepackage[colorlinks = true,
            linkcolor = blue,
            urlcolor  = black,
            citecolor = blue,
            anchorcolor = black]{hyperref}

\definecolor{goodorange}{RGB}{225,125,0}
\definecolor{goodgreen}{RGB}{5,130,5}
\definecolor{goodred}{RGB}{220,50,25}
\definecolor{goodblue}{RGB}{30,144,255}
\definecolor{OliveGreen}{RGB}{5,100,5}

\newcommand{\note}[2]{
\ifthenelse{\equal{#1}{F}}{
\colorbox{goodorange}{\textcolor{white}{\footnotesize \fontfamily{phv}\selectfont #1}}
    \textcolor{goodorange}{{\footnotesize \fontfamily{phv}\selectfont #2}}\xspace
}{}
\ifthenelse{\equal{#1}{R}}{
\colorbox{goodred}{\textcolor{white}{\footnotesize \fontfamily{phv}\selectfont #1}}
    \textcolor{goodred}{{\footnotesize \fontfamily{phv}\selectfont #2}}\xspace
}{}
\ifthenelse{\equal{#1}{N}}{
\colorbox{goodblue}{\textcolor{white}{\footnotesize \fontfamily{phv}\selectfont #1}}
    \textcolor{goodblue}{{\footnotesize \fontfamily{phv}\selectfont #2}}\xspace
}{}
\ifthenelse{\equal{#1}{M}}{
\colorbox{goodblue}{\textcolor{white}{\footnotesize \fontfamily{phv}\selectfont #1}}
    \textcolor{goodblue}{{\footnotesize \fontfamily{phv}\selectfont #2}}\xspace
}{}
}

\usepackage{titlesec}

\usepackage[fontsize=11pt]{scrextend}
\titleformat{\section}
{\normalfont\sffamily\bfseries\color{OliveGreen}}
{\thesection.}{0.25em}{\uppercase}

\titleformat{\subsection}[runin]
{\normalfont\sffamily\bfseries}
{\thesubsection}{0.25em}{}[.\;\;]

\titleformat{\subsubsection}[runin]
{\normalfont\sffamily\itshape}
{
}{0.25em}{}[.\;\;]

\titleformat{\suppinfo}
{\normalfont\sffamily\bfseries}
{\thesubsection}{0.25em}{}

\titlespacing*{\section}{0pt}{0.5\baselineskip}{0.01\baselineskip}
\titlespacing*{\subsection}{0pt}{0.125\baselineskip}{0.01\baselineskip}

\author{Nicholas H. Stair}
\affiliation{Department of Chemistry and Cherry Emerson Center for Scientific Computation, Emory University, Atlanta, Georgia, 30322, U.S.A.}

\author{Francesco A. Evangelista}
\email{francesco.evangelista@emory.edu}
\affiliation{Department of Chemistry and Cherry Emerson Center for Scientific Computation, Emory University, Atlanta, Georgia, 30322, U.S.A.}

\let\oldmaketitle\maketitle
\let\maketitle\relax

\title{QForte: an efficient state simulator and quantum algorithms library for molecular electronic structure}

\date{\today}

\begin{document}

\ifpreprint
\else
\twocolumn[
\begin{@twocolumnfalse}
\fi
\oldmaketitle

\begin{abstract}
We introduce a novel open-source software package \qft, a comprehensive development tool for new quantum simulation algorithms. \qft incorporates functionality for handling molecular Hamiltonians, fermionic encoding, ansatz construction, time evolution, and state-vector simulation, requiring only a classical electronic structure package as a dependency.
\qft also contains black-box implementations of a wide variety of quantum algorithms including (but not limited to): variational and projective quantum eigensolvers, adaptive eigensolvers, quantum imaginary time evolution, quantum Krylov methods, and quantum phase estimation.
We highlight two features of \qft: i) how the Python class structure of \qft enables the facile implementation of new algorithms, and ii) how existing algorithms can be executed in just a few lines of code.
\end{abstract}

\ifpreprint
\else
\end{@twocolumnfalse}
]
\fi

\ifpreprint
\else
\small
\fi

\noindent

\section{Introduction}
The past decade has seen tremendous progress in the development of quantum computational hardware, enabling early demonstrations of quantum advantage\cite{arute2019quantum}, and numerous non-trivial applications ranging from quantum simulation\cite{OMalley:2016dc, kandala2017hardware, colless2018computation, shen2017quantum, hempel2018quantum, nam2020ground} to constrained optimization.\cite{lin2016performance, wang2018quantum}
These advances have concurrently inspired rapid development of numerous quantum algorithms amenable to both noisy intermediate-scale quantum\cite{preskill2018quantum} (NISQ) and fault-tolerant devices.
In the field of quantum simulation of many-body systems, a variety of methods now exist\cite{Peruzzo:2014kca, McClean:2016bs,motta2019determining,mcardle2020quantum,bauer2020quantum} which vary dramatically in quantum resource requirements, often with a tradeoff between circuit depth and measurement overhead.
Unfortunately, new quantum algorithmic developments are rarely accompanied by detailed numerical comparison to existing algorithms.
The infrequency of cross comparison is largely due to the lack of widely accessible reference implementations of new quantum algorithms.
To help ameliorate this issue, we introduce a new open-source software project, \qft, the first black-box quantum algorithms library for molecular electronic structure.

\begin{figure}[h]
   \centering
   \includegraphics[width=3.25in]{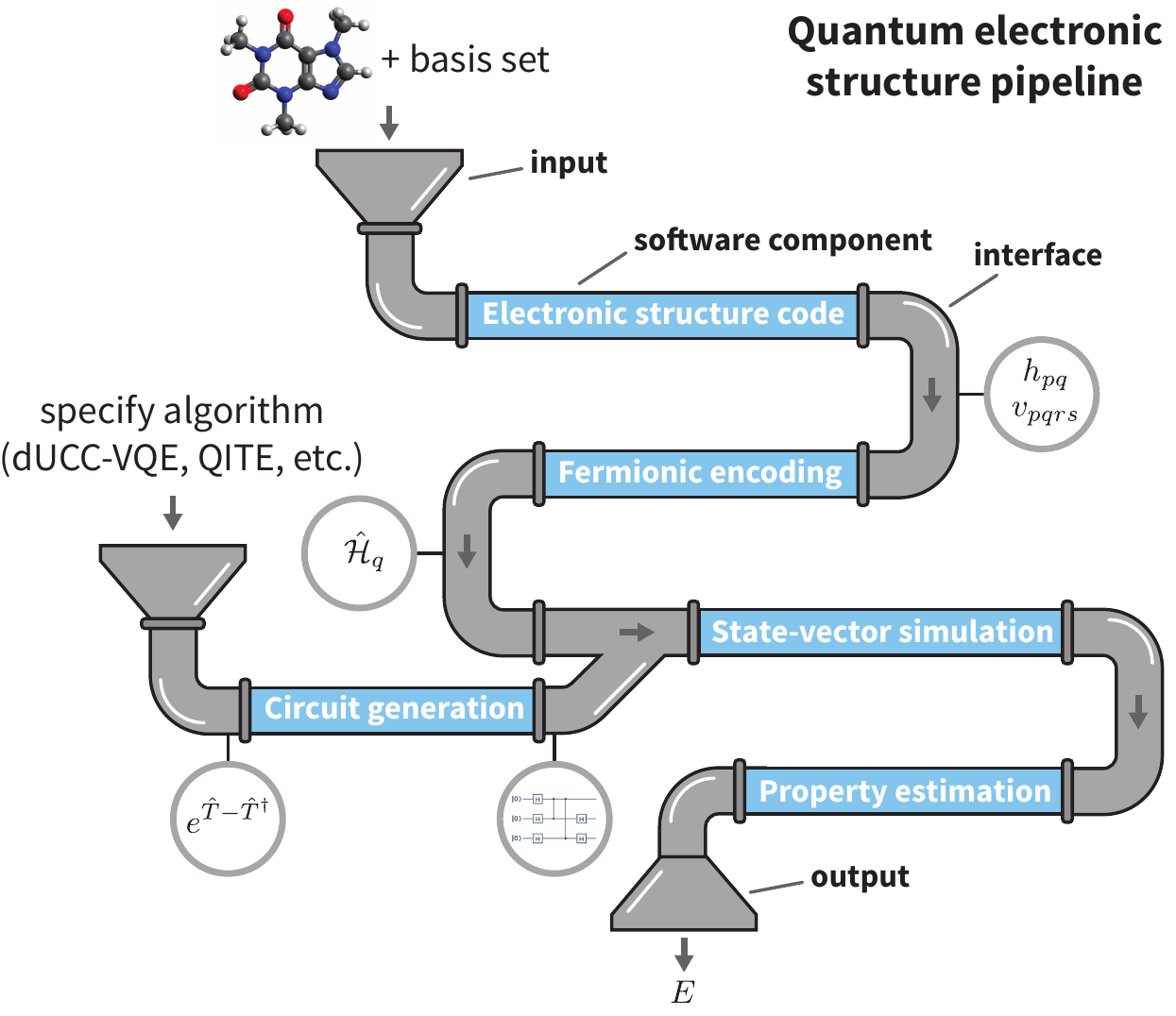}
   \caption{The quantum electronic structure pipeline.}
   \label{fig:quantum_pipeline}
\end{figure}

The software ecosystem for the development of new quantum algorithms is ever-expanding,
ranging from full-stack industry backed packages\cite{abraham2019qiskit, cirq_developers_2021_4750446, qsharp2020, smith2016practical, killoran2019strawberry}, to task specific open source projects.\cite{rubin2021fermionic, mcclean2020openfermion, smelyanskiy2016qhipster, luo2020yao, suzuki2020qulacs, kottmann2021tequila, bergholm2018pennylane, steiger2018projectq}
This is particularly true in the context of quantum algorithms for molecular electronic structure, because there are many complex software challenges involved in the complete user workflow pipeline illustrated in Fig.~\ref{fig:quantum_pipeline}, beginning with specification of a molecular geometry and ending with a numerical prediction of the molecular energy (and properties).
In order to accomplish this task, one must use a classical electronic structure package (e.g., \textsc{PySCF},\cite{sun2018pyscf} \textsc{Psi4},\cite{smith2020psi4} etc.) to obtain the matrix representation of one- ($h_{pq}$) and two-body ($v_{pqrs}$) operators (integrals) that define the molecular Hamiltonian
\begin{equation}
\hat{\mathcal{H}} = \sum_{pq} h_{pq} \cop{p} \aop{q} + \frac{1}{4} \sum_{pqrs} v_{pqrs} \cop{p} \cop{q} \aop{s} \aop{r}
\end{equation}
where $\cop{}$ and $\aop{}$ are, respectively, fermionic creation and annihilation operators labeled by the index of the spin orbital on which they act.
Next, one must utilize a package for fermionic encoding (e.g., \textsc{OpenFermion}), as well as appropriate application programming interfaces (API) (e.g.,\textsc{OpenFermion-PySCF} or \textsc{OpenFermion-Psi4}), to map fermionic operators (such as the Hamiltonian, $\hat{\mathcal{H}}$) to the so-called qubit representation ($\hat{\mathcal{H}}_q$) given as a linear combination of a given number ($N_\mathrm{PS}$) of Pauli strings (products of Pauli operators) $\hat{P}_\ell$:
\begin{equation}
\label{eq:qubit_hamiltonian}
\hat{\mathcal{H}} \xrightarrow{\substack{\text{fermionic} \\ \text{encoding}}} \hat{\mathcal{H}}_q = \sum_\ell^{N_\mathrm{PS}} h_\ell \hat{P}_\ell
\end{equation}
In order to apply quantum circuits associated with encoded operators (such as the time evolution unitary, or various unitary ans\"{a}tze)
one is required to install one of the numerous backend quantum-computer simulators such as those implemented in \textsc{Qiskit}\cite{abraham2019qiskit} (IBM), \textsc{CIRQ}\cite{cirq_developers_2021_4750446}  and \textsc{FQE}\cite{rubin2021fermionic} (Google), \textsc{Q\#}\cite{qsharp2020} (Microsoft), or \textsc{PyQuill}\cite{smith2016practical} (Rigetti), each associated with a distinct API.

In the outermost software layer exist packages such as \textsc{Tequila}\cite{kottmann2021tequila} that serve as sandbox implantation tools which solve many of the interoperability challenges associated with interfacing the aforementioned dependencies.
While flexible packages for sand-box development are undoubtably important, it is still usually left to the user to implement a desired algorithm, which is generally a non-trivial task given the diversity of quantum algorithms present in modern literature and the challenge of utilizing an inhomogeneous software ecosystem.
A single-package (incorporating all of the described steps) black-box implementation of quantum algorithms with which one can easily specify a molecular geometry and desired quantum algorithm (as is the case for classical electronic structure packages) is highly desirable for researchers interested in generating comparative results rapidly.

An additional challenge to black-box implementation of quantum algorithms is their now significant level of diversity.
Arguably the simplest and most well established (hybrid) algorithm is the variational quantum eigensolver\cite{Peruzzo:2014kca, McClean:2016bs} (VQE).
In VQE, the ground state is approximated by a trial state optimized in a procedure that combines a classical optimization algorithm with quantum measurement of the energy and gradients of the trial state.
While implementation of a vanilla VQE code is straightforward for toy examples, automatic generation of the ansatz circuit is generally more complicated and obviously dependent on the specific form of the ansatz (i.e. disentangled unitary coupled cluster,\cite{McClean:2016bs, barkoutsos2018quantum, Romero:2019hk} hardware-efficient,\cite{kandala2017hardware} Hamiltonian variational,\cite{wecker2015progress} qubit coupled cluster,\cite{Ryabinkin:2018jw} to name a few).
Moreover, many hybrid approaches such as adaptive ansatz approaches\cite{grimsley2019adaptive, ryabinkin2020iterative} or subspace expansion methods\cite{mcclean2017hybrid} incorporate the basic VQE schema as a subroutine and require additional implementation for determination of matrix elements to solve a generalized eigenvalue problem and/or an additional algorithmic layer to extend the ansatz.
Similar implementation challenges exist for algorithms that rely on (often controlled) Hamiltonian time evolution such as quantum phase estimation\cite{kitaev1995quantum, Abrams:1997ha, Abrams:1999ur} (QPE), or time-evolved subspace methods.\cite{Parrish:2019tc, Stair_2020, klymko2021real}
For algorithms that measure projected quantities such as quantum imaginary time evolution\cite{motta2019determining} (QITE) and the projective quantum eigensolver\cite{stair2021simulating} (PQE), one must additionally implement the (often iterative) parameter update procedure.

Our new open-source package \qft is an end-to-end electronic structure package for quantum algorithms, and is still capable of facilitating sand-box implementations of new algorithms, only relying on a classical electronic structure package as a dependency.
The remainder of this article is organized as follows.
In Sec.~\ref{sec:structoverview} we will describe key component classes in \qft and its interface to \textsc{Psi4}.
In Sec.~\ref{sec:qf_algs} will discuss each of the quantum algorithms currently implemented in \qft as well as some of their implementation details in terms of the key components.
In Sec.~\ref{sec:timings} we discuss representative timings for critical subroutines such as determining Hamiltonian expectation values.
Finally, in Sec.~\ref{sec:ex_usage} we demonstrate an example of how \qft can (i) be used to implement a new quantum algorithm and (ii) be used to compare new algorithms to the ones already implemented in its library.

\section{Overview of the structure of \qft}
\label{sec:qf_struct}

\begin{figure*}[htbp]
   \centering
   \includegraphics[width=5.5in]{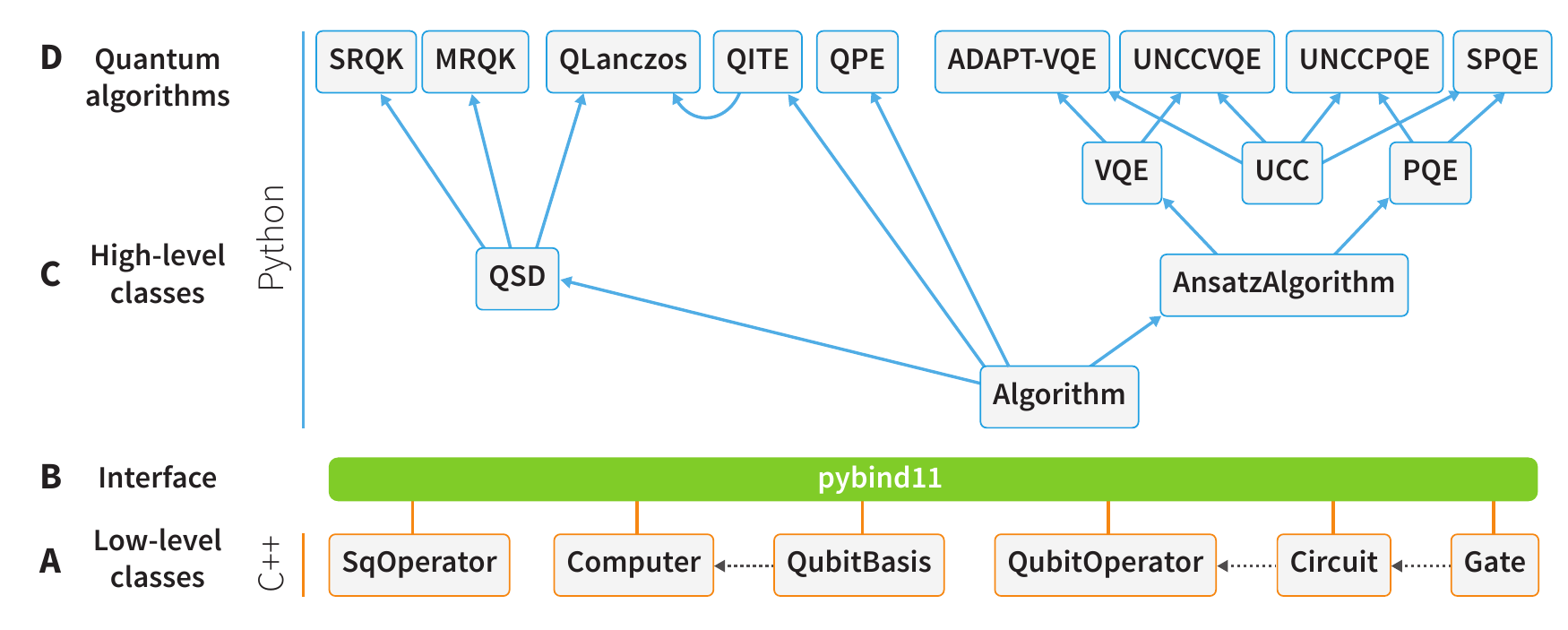}   \caption{The structure of the QForte package. Solid lines indicate inheritance (pointing from the base to the derived class) while dashed lines indicate use of objects from another class (pointing from the class used to the one using it). A) Low-level C++ classes are used to store and manipulate a state vector. B) The C++ classes are exposed to Python via the \textsc{pybind11} library. C) A high-level Python classes implement components of quantum algorithms. D) Quantum algorithms available to the user are implemented using classes from the high-level API.}
   \label{fig:qforte_structure}
\end{figure*}

\label{sec:structoverview}

In order to facilitate simple molecular-geometry to quantum algorithm energy functionality, \qft wraps the entire quantum electronic structure pipeline in a black-box code.
The main software components of \qft are illustrated in Fig.~\ref{fig:qforte_structure}.
The lowest level contains components that require efficient execution, including the state-vector simulator (\verb+Computer+) and quantum circuits (\verb+Circuit+).
These are implemented as classes in C++ and exposed in Python via \pybind.\cite{jakob2017pybind11}
Higher-level components such as the \verb+SystemFactory+ class which interfaces \qft to classical electronic structure packages, and the subclasses that implement algorithms are all written in Python and use the \pybind interface to the lower-level components.
Here we will give an overview of some of the most important components of \qft.

\subsection{The state-vector simulator}
An important aspect of \qft that distinguishes it from many other packages is its incorporation of a dedicated state-vector simulator.
State-vector simulators store and manipulate a classical representation of the full quantum state that exists on quantum hardware.

\subsubsection{The QubitBasis class}

The state-vector simulator in \qft makes heavy use of an elementary C++ \verb+QubitBasis+ class used to represent an element of a $n_\mathrm{qb}$ qubit Fock space basis  {$\mathcal{F} = \{ \ket{q_0} \otimes \cdots \otimes \ket{q_{n_\mathrm{qb}-1}} \}$.
Each instance of \verb+QubitBasis+ represents an element of $\mathcal{F}$, and is characterized by a 64-bit unsigned integer such that each bit represents the binary state of a qubit:
\begin{equation}
\underbrace{q_{63} \cdots q_1q_0}_{\text{64-bit unsigned integer}}
\equiv
\underbrace{\ket{q_0} \otimes \cdots \otimes \ket{q_{63}},}_{\text{element of the Fock-space basis}}
\text{ with } q_i \in \{0,1\}
\end{equation}
Because application of quantum gates to any basis element results in flipping the state of target qubit(s), use of the unsigned integer type is beneficial as it allows for very efficient bitwise operations.
The binary number corresponding to an element of the Fock-space basis also offers a convenient way to map the qubit multi-index $q_0q_1\cdots$ to a single index (address).

\subsubsection{The QuantumComputer class}
The backbone of the state-vector simulator in \qft is the \verb+Computer+ class, which, for a given number of qubits ($n_\mathrm{qb}$) stores a state vector of the form
\begin{equation}
\ket{\Psi_q}
= \sum_{q_0 \cdots q_{n_\mathrm{qb}-1}}^{\{0,1\}} C_{q_0,\ldots,q_{n_\mathrm{qb}-1}} \ket{q_0} \otimes \cdots \otimes \ket{q_{n_\mathrm{qb}-1}}
\end{equation}
By default, a \verb+Computer+ object is initialized in the state $\ket{\Psi_q} = \ket{0} \otimes \cdots \otimes \ket{0}$.
The \verb+Computer+ class is comprised of a
complex vector to store $C_{q_0,\ldots,q_{n_\mathrm{qb}-1}}$, as well as a vector of \verb+QubitBasis+ objects (both of dimension $2^{n_\mathrm{qb}}$ where $n_\mathrm{qb}$ is the number of qubits).
An example of how to instantiate a \verb+Computer+ with four qubits is shown in Lst.~\ref{lst:inst_qc}.
\begin{lstlisting}[language=Python, caption={Initializing a {\ttfamily Computer} object with four qubits.}, label={lst:inst_qc}]
import qforte as qf
nqb = 4
qcomp = qf.Computer(nqb)
\end{lstlisting}

\subsubsection{The \code{Gate} class}
Once a \verb+Computer+ is initialized, the state can be modified by applying gates, encoded in the class \code{Gate}.
The \code{Gate} class (when used in conjunction with a \code{Computer}) is the most fundamental building block for all quantum algorithms in \qft.
Some of the most pertinent gates used in quantum simulation are the Pauli gates ($X$, $Y$, and $Z$), the Hadamard gate $H$ (not to be confused with the Hamiltonian $\hat{\mathcal{H}}$), the controlled NOT [CNOT] gate, and the parametric z rotation gate $R_z(\theta)$. A full list of gates implemented in \qft can be found in the documentation.

The \code{Gate} class has several important attributes including its type,
the target (and optionally the control) qubit index, and a matrix of complex values that represents the operator.
Instantiating a \code{Gate} is simply done via the \verb+gate()+ member function, as shown in Lst.~\ref{lst:inst_gate}.
\begin{lstlisting}[language=Python, caption={Initializing a {\ttfamily Gate} object. Here we instantiate a Pauli $X$ gate that will target the qubit $q_4$.}, label={lst:inst_gate}]
target_idx = 4
X_4gate = qf.gate('X', target_idx)
\end{lstlisting}

Listing~\ref{lst:bell_state} shows a small example of using \qft's state-vector simulator to construct the two-qubit Bell state
\begin{equation}
\label{eq:bell}
\ket{\Psi_\textrm{Bell}} = \frac{1}{\sqrt{2}} \ket{00} + \frac{1}{\sqrt{2}} \ket{11}
\end{equation}
by applying $H_0$ followed by ${\rm CNOT}_{0,1}$ to the state $\ket{00}$.
Recall that the action of the Hadamard gate $H$ is:
\begin{equation}
H\ket{0} = \frac{1}{\sqrt{2}} \big( \ket{0} + \ket{1} \big), \quad
H\ket{1} = \frac{1}{\sqrt{2}} \big( \ket{0} - \ket{1} \big)
\end{equation}
Recall that the action of the controlled NOT gate [with target qubit $q_0$, and control qubit $q_1$ (${\rm CNOT}_{0,1}$)] is:
\begin{equation}
\begin{split}
{\rm CNOT}_{0,1}\ket{00} = \ket{00}\quad
{\rm CNOT}_{0,1}\ket{01} = \ket{11}\\
{\rm CNOT}_{0,1}\ket{10} = \ket{10}\quad
{\rm CNOT}_{0,1}\ket{11} = \ket{01}
\end{split}
\end{equation}

\begin{lstlisting}[language=Python, caption={Creating a Bell state using a circuit that applies the unitary ${\rm CNOT}_{0,1} H_0$ to the state $ \ket{00}$.}, label={lst:bell_state}]
# Initialize a two-qubit QuantumComputer.
nqb = 2
qbell = qf.Computer(nqb)

# Initialize the gates needed to build the Bell state.
H_0 = qf.gate('H', 0)
CNOT_0_1 = qf.gate('CNOT', 1, 0)

# Apply the Hadamard gate.
qbell.apply_gate(H_0)

# Apply the CNOT gate.
qbell.apply_gate(CNOT_0_1)
\end{lstlisting}

In \qft, the action of a gate on a quantum state (represented by a \code{Computer} object) is implemented with efficient algorithms specialized for different gates.
For example, in Fig.~\ref{fig:x_gate} we illustrate how the $X_2$ operator is applied to a 3-qubit state.
Since $X_2$ modifies only the first bit (by flipping it), the operation $X_2 \ket{\Psi} \rightarrow \ket{\Psi'}$ can be efficiently implemented with a low-level copy operation that realizes the following action on the coefficients in a \code{Computer} object: $C'_{q_0q_1q_2} = C_{q_0q_1(1- q_2)}$.
Since the operation is performed on continuous sections of the state vector, the effect of cache misses is minimized and the operation can be easily vectorized on multi-core architectures. 
The same principle illustrated here can be applied to more complex one-qubit gates such as parameterized rotations and two-qubit gates such as $\rm{CNOT}$.

\begin{figure}[h]
   \centering
   \includegraphics[width=3.25in]{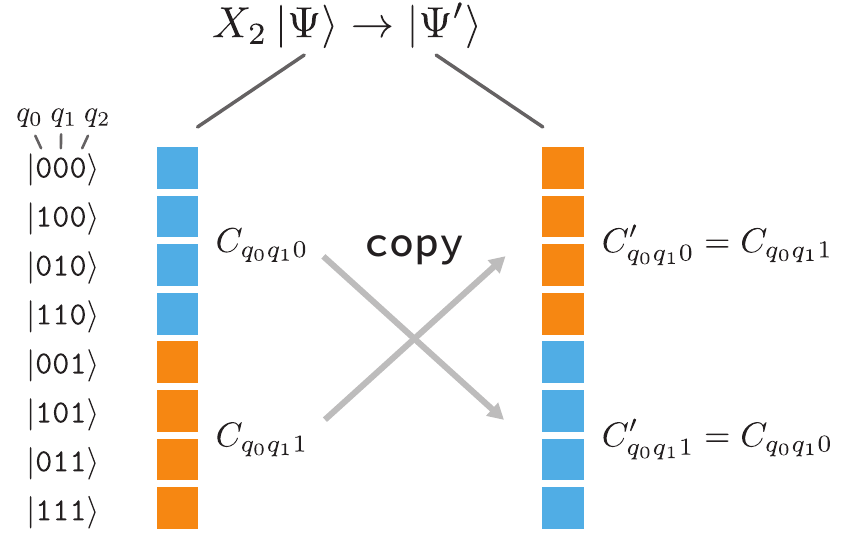}
   \caption{Example of how gate operations are implemented in \qft.
   In this example, the $X_2$ gate is applied to a general state $\Psi$ producing a new state $\Psi'$. The action of $X_2$ can be implemented efficiently with a low-level C++ copy operation to two separate portions of the state vector.}
   \label{fig:x_gate}
\end{figure}

\subsubsection{The Circuit class}
In virtually any quantum algorithm it is necessary to apply many gates sequentially.
A so-called quantum circuit, commonly referred to as a unitary ($\hat{U}$), is represented by a product of quantum gates, making the overall circuit itself a unitary operation.
The \verb+Circuit+ class operates at one level above the \code{Gate} class and its primary attribute is a vector of \code{Gate} objects.

Although any product of elementary gates technically constitutes a circuit, one of the most important circuit structures in quantum simulation is that which represents unitaries obtained by exponentiating a product of Pauli operators:
\begin{equation}
\label{eq:exp_pt2}
e^{i \theta_\ell \hat{P}_\ell}
\end{equation}
where
\begin{equation}
\hat{P}_\ell = \prod_k^{n_\ell} \hat{\sigma}^{(\ell)}_k
\end{equation}
is a unique product of $n_\ell$ Pauli operators $\hat{\sigma}^{(\ell)}_k$.
The compound index $k=(p, [X, Y, \text{or } Z])$ labels a combination of a specific Pauli operator and the qubit ($p$) on which it acts.
The function \verb+exponentiate_pauli_string+ in \qft is responsible for converting Eq.~\eqref{eq:exp_pt2} into a circuit containing one- and two-qubit gates.
An example of how one would construct such a circuit in \qft for the operator $\exp(-i 0.5 X_2 Z_1 Z_0)$ is shown in Lst.~\ref{lst:exp_circ}, while in Fig.~\ref{fig:exp_pt} we report the corresponding quantum circuit generated by the function \verb+exponentiate_pauli_string+.
This algorithm follows a standard approach\cite{Peruzzo:2014kca, yung2014transistor, McClean:2016bs} of using operator identities (e.g., like $X = HZH$) to express the starting expression in terms of the exponential of products of $Z$ operators only, namely
\begin{equation}
\begin{split}
\exp(-i 0.5 \, X_2 Z_1 Z_0)
&= \exp(-i 0.5 \, H_2 Z_2 H_2 Z_1 Z_0) \\
&= H_2 \exp(-i 0.5 \, Z_2 Z_1 Z_0)  H_2
\end{split}
\end{equation}
where one uses the fact that $H_2$ is its own inverse ($H_2 H_2 = 1$) and that it commutes with $Z_0$ and $Z_1$.
Next, the term $\exp(-i 0.5 \, Z_2 Z_1 Z_0)$, responsible for the fermionic sign, is implemented as a cascade of CNOT gates, a z-axis rotation of $2\theta_\ell$, and the inverse of the CNOT cascade:
\begin{equation}
\begin{split}
e^{-i 0.5 \, Z_2 Z_1 Z_0} = {\rm CNOT}_{0,1} {\rm CNOT}_{1,2} R_z(1.0) {\rm CNOT}_{1,2} {\rm CNOT}_{0,1}
\end{split}
\end{equation}

\begin{lstlisting}[language=Python, caption={Constructing the circuit corresponding to the operator $\exp(-i 0.5 X_2 Z_1 Z_0)$.}, label={lst:exp_circ}]
# Construct the desired preliminary circuit (X2 Z1 Z0)
circ = qf.Circuit()
circ.add_gate(qf.gate('Z', 0))
circ.add_gate(qf.gate('Z', 1))
circ.add_gate(qf.gate('X', 2))

# Define the factor = -i theta
factor = -0.5j

# Construct the unitary for the exponential.
Uexp, phase = exponentiate_pauli_string(factor, circ)
\end{lstlisting}

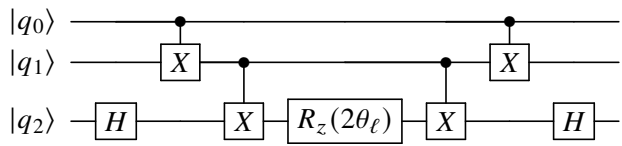
\begin{figure}
\centering
\[ \Qcircuit @C=0.85em @R=.65em {
   & \lstick{\ket{q_{0}}} &\qw	& \ctrl{1} \qw &  \qw &  \qw  &  \qw  & \ctrl{1} \qw & \qw     &\qw \\
   & \lstick{\ket{q_{1} }} &\qw &  \gate{X}  &  \ctrl{1} \qw     &  \qw &  \ctrl{1} \qw      & \gate{X}          & \qw    & \qw\\
   & \lstick{\ket{q_{2} }} &\gate{H} &  \qw &  \gate{X} &  \gate{R_z(2\theta_\ell)}                                        &  \gate{X}  & \qw & \gate{H} & \qw
} \]
\caption{Circuit diagram for exponential unitary operators commonly used in many quantum simulation algorithms. This example considers operators of the form $\exp(-i \theta_\ell X_2 Z_1 Z_0)$.}
\label{fig:exp_pt}
\end{figure}

\subsection{The QubitOperator and SQOperator classes}
The outer-most operations class in QForte is the \verb+QubitOperator+ class.
This class represents operators $\hat{O}$ that are linear combinations of $N_\ell$ unitaries ($\hat{U}_\ell$) as
\begin{equation}
\hat{O} = \sum_\ell u_\ell \hat{U}_\ell
\end{equation}
where $u_\ell$ is a complex coefficient.
The key attribute of the \verb+QubitOperator+ class is a vector of pairs containing a complex coefficient and a \verb+Circuit+ object.
In \qft's implementation, the \verb+QubitOperator+ class is used to represent objects such as the Hamiltonian $\hat{\mathcal{H}_q}$ or the cluster operator\cite{crawford2000introduction} $\hat{T}_q$ (in their qubit representations).
It is important to note that although it is always possible in \qft to apply a \verb+QubitOperator+ to a state, unless $\hat{O}$ is unitary, this operation cannot be realized physically on a quantum device.

The ability to evaluate expectation values of the form $\expval{\hat{O}} =  \bra{\Psi_q} \hat{O} \ket{\Psi_q}$ is also paramount to many quantum algorithms.
Examples include the energy and energy gradients for VQE, as well as projected quantities such as off-diagonal matrix elements in QSD techniques, or residuals in PQE.
In \qft this is accomplished easily by passing the operator to an exception value function. Listing~\ref{lst:exp_val} for example shows how one would measure the expectation value of the operator $\hat{O} = X_0 + \hat{X_1}$ with respect to the Bell state (constructed in Lst.~\ref{lst:bell_state}). 
We note that it is also possible to determine approximate expectation values in \qft based on a user specified number of measurements.

\begin{lstlisting}[language=Python, caption={Measuring expectation values for  {\ttfamily QubitOperators}. }, label={lst:exp_val}]
# Initialize the operator
O = QubitOperator()
O.add_term(1.0, build_circuit('X_0'))
O.add_term(1.0, build_circuit('X_1'))

# Get the expectation value.
exp_val = qbell.direct_op_exp_val(O)
\end{lstlisting}

\qft also supports operators in the form of second quantization, that is, operators comprised of fermionic annihilation and creation operators.
The \verb+SQOperator+ class functions very similarly to the \verb+QubitOperator+ class, but utilizes a slightly different syntax.
We note that second quantized operators in \qft always assume that the individual fermionic operators are normal ordered (creation operators appear to the left of annihilation operators) within a term.
The second quantized operators can then be transformed to the \verb+QubitOperator+ representation (given as a linear combination of products of Pauli operators) via the Jordan-Wigner (JW) transformation.\cite{jordan1993paulische} Other common fermionic encodings\cite{bravyi2002fermionic, havlivcek2017operator, setia2018bravyi, setia2019superfast} are not implemented natively in \qft and will be added in future releases.
Under the JW transformation, there is a one-to-one mapping between a spin orbital $\phi_p$ and qubit $q_p$ such that the fermionic annihilation ($\hat{a}_{p}$) and creation ($\hat{a}^{\dagger}_{p}$) operators are represented by combinations of Pauli strings
\begin{equation}
\hat{a}_{p} = \frac{1}{2} \Big( X_p + i Y_p \Big) Z_{p-1} \dots Z_0
\end{equation}
and,
\begin{equation}
\hat{a}^{\dagger}_{p} = \frac{1}{2} \Big( X_p - i Y_p \Big) Z_{p-1} \dots Z_0
\end{equation}
Listing~\ref{lst:init_sqop} shows how to instantiate a \verb+SQOperator+ with the operator $0.5 \, \cop{1}\aop{2} - 0.25 \, \cop{4}\cop{2}\aop{3}\aop{1} $ and transform it into a \verb+QubitOperator+.
\begin{lstlisting}[language=Python, caption={Converting {\ttfamily SQOperators} into {\ttfamily QuantumOperators}. This example constructs the operator $0.5 \, \cop{1}\aop{2} - 0.25 \, \cop{4}\cop{2}\aop{3}\aop{1}$.}, label={lst:init_sqop}]
# Initialize the second quantized operator.
sq_op = qf.SQOperator()

# Construct the terms and add them to the list.
h1 = 0.5
h2 = -0.25j
sq_op.add_term(h1, [1], [2])
sq_op.add_term(h2, [4, 2], [3, 1])

# Transform to the qubit operator representation.
pauli_op = sq_op.jw_transform()
\end{lstlisting}

\subsection{The molecule class}
\label{sec:mol_class}
As discussed in the introduction, the first step in an quantum electronic structure computation is to obtain the molecular Hamiltonian in the qubit operator representation, based on a specified molecular geometry.
In \qft this is all accomplished using the \verb+system_factory+ and \verb+molecule+ classes.
To begin, as shown in Lst.~\ref{lst:inst_mol}, one simply imports the appropriate modules and specifies a geometry (in this case as a list of element symbols and xyz coordinates).
Once the molecule class has been populated, the user has access to the molecular Hamiltonian both in its second-quantization representation (as a \verb+SQOperator+) and in a qubit representation (as a \verb+QubitOperator+) resulting from the Jordan-Wigner transformation.
The molecule object is a key data structure in \qft that is passed to all algorithms to perform a quantum computation.

\begin{lstlisting}[language=Python, caption={Initializing the \qft molecule object.}, label={lst:inst_mol}]
from qforte import *

# Define the molecular geometry.
geom = [('H', (0., 0., 0.)), ('H', (0., 0., 0.75))]

# Instantiate the system_factory object (also populates the integrals).
factory = system_factory(build_type='psi4', mol_geometry=geom, basis='sto-3g')

# Get the molecule object.
H2mol = factory.get_molecule()
\end{lstlisting}

\section{Algorithms implemented in \qft}
\label{sec:qf_algs}

Quantum algorithms are implemented in \qft in the Python layer, and are built by composing abstract base classes (ABCs) that implement algorithms (and optionally) mixin classes that encode specific ans\"{a}tze.
For example, the lowest-level ABC in \qft is \code{Algorithm}, which is responsible for defining attributes and abstract functions common to all derived classes.
The class \code{ AnsatzAlgorithm}, derived from \code{Algorithm}, defines additional functionality that builds a quantum circuit for any algorithm that employs a parameterized ansatz.  
Each distinct algorithm implemented in \qft is then defined by its own concrete class that inherits from various appropriate parent classes, as displayed in Fig.~\ref{fig:qforte_structure}.

We consider the algorithms implemented in \qft partitioned into several overlapping categories.
The first is variational hybrid algorithms\cite{Peruzzo:2014kca, McClean:2016bs, cerezo2020variational} in which a classical optimizer is utilized to minimize the energy expectation value.
The second category is projective approaches\cite{motta2019determining, stair2021simulating} where projective quantities are measured on a quantum device and then used to directly update or augment a classically parameterized unitary.
The third category is quantum subspace diagonalization\cite{mcclean2017hybrid, motta2019determining, Parrish:2019tc, huggins2020non, Stair_2020} (QSD) in which a non-orthogonal many body basis is generated from a family of operators, and the matrix elements of a generalized eigenvalue problem are measured on a quantum device.
Finally, there are those algorithms derived from quantum phase estimation\cite{Abrams:1997ha, Abrams:1999ur} where the eigenvalue of a time evolved state is estimated by a binary readout of a set of ancilla qubits.

Currently, \qft contains black-box implementations of the following algorithms: (i) variational quantum eigensolver\cite{Peruzzo:2014kca, McClean:2016bs} (VQE) and (ii) projective quantum eigensolver\cite{stair2021simulating} (PQE) both with a disentangled\cite{evangelista2019exact} (factorized) unitary coupled cluster\cite{szalay1995alternative, taube2006new, cooper2010benchmark, evangelista2011alternative, harsha2018difference} (dUCC) ansatz, (iii) the adaptive derivative-assembled pseudo-trotter (ADAPT)-VQE\cite{grimsley2019adaptive}, (iv)  selected PQE\cite{stair2021simulating} (SPQE), (v) a variant of quantum imaginary time evolution\cite{motta2019determining} (QITE) and its quantum Lanczos\cite{motta2019determining} (QLanczos) extension, (vi) quantum Krylov\cite{Parrish:2019tc, Stair_2020} (QK) and (vii) its selected multireference variant\cite{Stair_2020} (MRSQK), and a pilot implementation of (vii) quantum phase estimation\cite{Abrams:1997ha, Abrams:1999ur} (QPE).
Each of these algorithms is implemented using the software components described in Sec.~\ref{sec:structoverview}.
In the following subsections we will briefly summarize each of these methodologies.

\subsection{Variational quantum eigensolver (VQE) based on dUCC trial states}

In the general VQE schema, one uses a unitary circuit $\hat{U}(\bm{\theta})$ parameterized by the vector $\bm{\theta}$ to construct a normalized trial state of the form
\begin{equation}
\ket{\Psi_\mathrm{VQE}} = \hat{U}(\bm{\theta}) \ket{\Phi_0}
\end{equation}
from an easily prepared reference state $\ket{\Phi_0}$ (such as the Hartree--Fock state).
One then minimizes the energy expectation value of the trial state
\begin{equation}
E(\bm{\theta}) = \bra{\Phi} \hat{U}^\dagger(\bm{\theta}) \hat{\mathcal{H}} \hat{U}(\bm{\theta}) \ket{\Phi}
\end{equation}
This minimization problem is solved by a classical optimization algorithm that calls a quantum routine to access the energy (and optionally  gradients\cite{schuld2019evaluating, kottmann2020feasible}) of the trial state.
Energy and gradients may be computed by averaging the expectation value of each term in the qubit Hamiltonian [Eq.~\eqref{eq:qubit_hamiltonian}], for example
\begin{equation}
E(\bm{\theta})  = \sum_\ell h_\ell \bra{\Phi} \hat{U}^\dagger(\bm{\theta}) \hat{P}_\ell\hat{U}(\bm{\theta}) \ket{\Phi}
\end{equation}

In \qft we have implemented VQE with a disentangled (or factorized) UCC ansatz $\hat{U}_\mathrm{dUCC}(\mathbf{t})$.
We assume a single determinant reference state $\ket{\Phi_0} = \ket{\phi_1 \phi_2 \cdots}$ specified by occupied spin orbitals $\{ \phi_i \}$ and unoccupied (virtual) spin orbitals $\{ \phi_a \}$.
The operator $ \hat{\tau}_\mu \equiv  \hat{\tau}_{ij\cdots}^{ab\cdots} = \cop{a} \cop{b} \cdots \aop{j} \aop{i}$ is a particle-hole excitation operator that transforms the reference determinant $\ket{\Phi_0}$ into the excited determinant $\ket{\Phi_\mu} = \hat{\tau}_\mu\ket{\Phi_0}$.
The dUCC ansatz is then constructed as a product of exponentiated anti-hermitian operators $\hat{\kappa}_\mu \equiv \hat{\tau}_\mu - \hat{\tau}_\mu^\dagger$ as
\begin{equation}
\label{eq:ducc}
\hat{U}_\mathrm{dUCC}(\mathbf{t})=
\prod_\mu e^{ t_\mu \hat{\kappa}_\mu}
\end{equation}
The dUCC circuit is built automatically in \qft by initializing a \verb+SQOperator+ list that represent the $\hat{\kappa}_\mu$, transforming to a \code{QuantumOperator} list via the Jordan--Wigner transformation, and constructing the circuit for the exponential of each sub-term [Eq.~\eqref{eq:exp_pt2}].
We note that optimization in \qft relies on the \textsc{SciPy} scientific computing library,\cite{virtanen2020scipy} and that any string specifying a valid \textsc{SciPy} optimizer may be passed to the \code{run()} function of a VQE algorithm.
The ordering of the operators $e^{t_{\mu} \hat{\kappa}_{\mu}}$ entering Eq.~\eqref{eq:ducc} is defined by the binary representation of the corresponding determinants $\ket{\Phi_{\mu}} = \hat{\tau}_{\mu} \ket{\Phi_\textrm{HF}}$ in the occupation number representation.
The maximum excitation level (single and double, triple, etc\dots) for the operators in Eq.~\eqref{eq:ducc} can be of arbitrary rank and is specified by the user as an option passed to the \code{run()} function.
\subsection{Adaptive VQE}

We have also implemented the ADAPT-VQE approach, and adaptive variant of dUCC-VQE.\cite{grimsley2019adaptive}
In ADAPT-VQE, the unitary ansatz at macro-iteration $k$ is defined as
\begin{equation}
\hat{U}_\mathrm{ADAPT}^{(k)}(\mathbf{t}) = \prod_\nu^{k} e^{ t_\nu^{(k)} \hat{\kappa}_\nu^{(k)} }
\end{equation}
where $\nu$ is a compound index corresponding to operators $\hat{\kappa}_\nu$ in a pool $\mathcal{P}$. In \qft it is possible to construct $\mathcal{P}$ in a variety of ways, including the generalized single and double excitation/de-excitation operators described in the original publication.\cite{grimsley2019adaptive}
The parameters $t_\nu^{(k)}$ are optimized at each macro-iteration employing the general VQE schema.
New operators are determined from the pool by computing the energy gradient
\begin{equation}
g_\nu = \bra{\Psi_\mathrm{VQE}} [ \hat{\mathcal{H}}, \hat{\kappa}_\nu ] \ket{\Psi_\mathrm{VQE}}
\end{equation}
with respect to $t_\nu$ of each operator in $\mathcal{P}$ and selecting the operator with the largest gradient magnitude. This new operator is placed at the end of the ansatz in the next iteration, and the procedure is terminated when the norm of the gradient vector for the pool falls below a user-provided threshold that controls the accuracy of the ansatz.

\subsection{Projective quantum eigensolver (PQE) based on dUCC trial states}
In the dUCC projective quantum eigensolver approach we consider a trial state of the form $\ket{\Psi} = \hat{U}(\bm{t}) \ket{\Phi_0}$, where $\hat{U}(\bm{t})$ is defined by Eq.~\eqref{eq:ducc}.
PQE aims to solve the following unitarily-transformed version of the Schr\"{o}dinger equation, obtained from the original one by left-multiplying with $\hat{U}^\dagger(\bm{t})$,
\begin{equation}
\hat{U}^\dagger(\mathbf{t}) \hat{\mathcal{H}} \hat{U}(\mathbf{t}) \ket{\Phi_0} = E \ket{\Phi_0}
\end{equation}
Rather than accomplishing this via variational minimization (as is done in VQE), PQE seeks to minimize the residual condition
\begin{equation}
r_\mu(\mathbf{t}) \equiv \bra{\Phi_\mu} \hat{U}^\dagger(\mathbf{t}) \hat{\mathcal{H}} \hat{U}(\mathbf{t}) \ket{\Phi_0} = 0,\text{ for all } \Phi_\mu \in Q
\label{eq:ucc2}
\end{equation}
where the residual $r_\mu(\bm{t})$ is given by the projection of the unitarily-transformed  Schr\"{o}dinger equation onto the excited determinant $\ket{\Phi_\mu}$.
In practice we only consider enforcing the residual condition for a subset $Q$ of excited determinants (such as all single and double excitations).

The residuals $r_\mu$ can be easily determined from symmetric expectation values and can therefore be measured via operator averaging on a quantum device.
Specifically, each element $r_\mu$ is obtained as
\begin{equation}
\begin{split}
r_\mu(\bm{t}) = &
 \bra{\Omega_\mu} \hat{U}^\dagger(\mathbf{t}) \hat{\mathcal{H}} \hat{U}(\mathbf{t}) \ket{\Omega_\mu} \\
 &-\frac{1}{2} \bra{\Phi_0} \hat{U}^\dagger(\mathbf{t}) \hat{\mathcal{H}} \hat{U}(\mathbf{t}) \ket{\Phi_0}\\
&-\frac{1}{2}\bra{\Phi_\mu} \hat{U}^\dagger(\mathbf{t}) \hat{\mathcal{H}} \hat{U}(\mathbf{t}) \ket{\Phi_\mu}
\end{split}
\end{equation}
where $\Omega_\mu$ is an easily preparable superposition of $\ket{\Phi_0}$ and $\ket{\Phi_\mu}$
\begin{equation}
\ket{\Omega_\mu} = e^{\frac{\pi}{4} \hat{\kappa}_\mu} \ket{\Phi_0} = \frac{1}{\sqrt{2}} \ket{\Phi_0} + \frac{1}{\sqrt{2}} \ket{\Phi_\mu}
\end{equation}
One of the most important features of dUCC-PQE is that the parameter vector $\bm{t}$ can be updated by measuring the residuals via a simple quasi-Newton iteration approach
\begin{equation}
\label{eq:fixed_point}
t_\mu^{(n +1)} = t_\mu^{(n)} + \frac{r^{(n)}_\mu}{\Delta_\mu}
\end{equation}
where the superscript ``$(n)$'' indicates the amplitude at iteration $n$.
The quantities $\Delta_\mu$ are standard M{\o}ller--Plesset denominators $\Delta_\mu \equiv \Delta_{ij\cdots}^{ab\cdots} = \epsilon_i + \epsilon_j + \ldots -\epsilon_a -\epsilon_b \ldots$ where $\epsilon_i$ are Hartree--Fock orbital energies.

\subsection{Selected PQE}

The selected ansatz variation of PQE (SPQE) is also implemented in \qft and, similarly to ADAPT, it utilizes a dUCC ansatz constructed iteratively from a (growing) set of operators $\mathcal{A}$.
In brief, the selection procedure is done by construction of a normalized state $\ket{\tilde{r}}$ defined as
\begin{equation}
    \begin{split}
        \ket{\tilde{r}} &= \hat{U}^\dagger (\bm{t}) e^{i \Delta t \hat{\mathcal{H}}} \hat{U}(\bm{t}) \ket{\Phi_0} \\
        &=  (1 + i\Delta t \hat{U}^\dagger(\bm{t}) \hat{\mathcal{H}} \hat{U}(\bm{t}))  \ket{\Phi_0} + \mathcal{O}(\Delta t^2)
    \end{split}
\end{equation}
for which the square moduli of its probability amplitudes $|C_\mu|^2 \approx (\Delta t)^2 |\langle\Phi_\mu| \hat{U}^\dagger(\bm{t}) \hat{\mathcal{H}} \hat{U}(\bm{t}) |\Phi_0\rangle|^2$ are proportional to residuals $r_\mu$.
In \qft the time step is taken as a parameter of the calculation and the Suzuki-Trotter approximation\cite{trotter1959product, suzuki1993improved} is used for the time evolution operator.
We may then approximate the values of the (normalized) squared residuals as
\begin{equation}
\label{eq:approx_res_sq}
|\tilde{r}_\mu|^2 \approx \frac{N_\mu }{M}
\end{equation}
where $N_\mu$ is the number of times the state $\ket{\Phi_\mu}$ is measured from $M$ preparations of $\ket{\tilde{r}}$.
A cumulative thresholding procedure is then utilized to add new operators $\hat{\kappa}_\mu$ (corresponding to $|\tilde{r}_\mu|^2$) to the ansatz, enforcing the condition
\begin{equation}
\sum_{\hat{\kappa}_\mu \notin \mathcal{A}}^\text{excluded} |r_\mu|^2 \approx \sum_{\hat{\kappa}_\mu \notin \mathcal{A}}^\text{excluded} \frac{ |\tilde{r}_\mu|^2}{\Delta t^2} \leq \Omega^2
\label{eq:op_thresh}
\end{equation}
where $\Omega$ is a user-specified convergence threshold parameter.
This selection strategy is particularly appealing for strongly correlated systems because it does not require the candidate operators $\hat{\kappa}_\mu$ to be restricted to any particular excitation order.

\subsection{Quantum imaginary time evolution}
The quantum imaginary time evolution (QITE) algorithm\cite{motta2019determining} obtains the ground state $\Psi_0$ by propagating a trial state $\ket{\Phi}$ (assuming $\braket{\Psi_0}{\Phi} \neq 0$) via imaginary time evolution $e^{-\beta \hat{\mathcal{H}} }$ in the limit of infinite propagation time,
\begin{equation}
\ket{\Psi_0} = \lim_{\beta \rightarrow \infty} \frac{1}{\sqrt{ N(\beta) }} e^{-\beta \hat{\mathcal{H}} } \ket{\Phi}
\end{equation}
The factor $1 / \sqrt{ N(\beta) } = |\bra{\Phi} e^{-2 \beta \hat{\mathcal{H}}} \ket{\Phi}|^{-1/2}$ guarantees normalization of the evolved state.
Since the imaginary time evolution operator is non-unitary, its direct implementation on quantum computers is impractical.
QITE circumvents this limitation by constructing a unitary that approximates the action of the imaginary time evolution operator for a small time step $\Delta\beta$.
This unitary is obtained by matching a state propagated from imaginary time $\beta$ to $\beta + \Delta \beta$ with the unitary $\exp(-i \Delta\beta \hat{A})$, where $\hat{A}$ is Hermitian:
\begin{equation}
\label{eq:qite}
 \ket{\psi(\beta + \Delta \beta)} = N(\Delta \beta)^{-1/2} e^{-\Delta \beta \hat{\mathcal{H}} } \ket{\psi(\beta)}  = e^{-i \Delta\beta \hat{A} } \ket{\psi(\beta)}
\end{equation}
The operator $\hat{A}$ can be written as a linear expansion of Pauli strings $\hat{ \rho}_\mu  = \prod_l \hat{\sigma}_{\mu_l}^{(l)}$ such that $\hat{A} = \sum_{\mu \in \mathcal{P}} \alpha_\mu \hat{\rho}_\mu$.
Here, $\mathcal{P}$ is a subset with dimension $M$ of all possible $4^{N_{\rm{qb}}}$ Pauli strings, $\mu \equiv (\mu_1, \mu_2, .., \mu_{N_{\rm{qb}}}) $ is a multi-index describing a unique Pauli operator product, and $\mu_l \in \{ I, X, Y, Z \}$.
Equations for QITE are obtained by expanding Eq.~\eqref{eq:qite} up to linear terms and left-projecting onto $\bra{\Phi}\hat{A}^\dagger$, yielding the $M$-dimensional linear system $\mathbf{S}\boldsymbol{\alpha} = \mathbf{b}$.
The elements of $\mathbf{S}$, and $\mathbf{b}$, as well as the value of $N(\Delta\beta)$ can be determined via measurement of simple symmetric expectation values\cite{motta2019determining}:
\begin{equation}
N(\Delta \beta) \approx 1 - 2 \Delta \beta \bra{\Phi} \hat{\mathcal{H}} \ket{\Phi}
\end{equation}
\begin{equation}
S_{\mu \nu} = \bra{\Phi} \hat{\rho}_\mu^\dagger \hat{\rho}_\nu \ket{\Phi}
\end{equation}
and,
\begin{equation}
b_{\mu} = \frac{-i}{\sqrt{N(\Delta \beta)}} \bra{\Phi} \hat{\rho}_\mu^\dagger \hat{\mathcal{H}} \ket{\Phi}
\end{equation}
Here $\ket{\Phi}$ is an approximation to the exact time evolved state $\ket{\psi(\beta)}$.
We note that the overall QITE procedure is iterative, and requires that the linear system be solved $M = \beta / \Delta\beta$ times to reach the target evolution time $\beta$.

In \qft, the Pauli strings that enter into $\mathcal{P}$ are generated automatically from a user-specified manifold of second-quantized operators. Specifically, given a set of excitation/de-excitation operators, these are first Jordan-Wigner transformed, and all unique terms containing an odd number of $Y$ gates are included in $\mathcal{P}$.
We note that the \qft implementation of QITE differs from the ``inexact QITE'' described in Ref.~\citenum{motta2019determining}, in which a subgroup of important Pauli operators is chosen for \textit{each} $k$-local Hamiltonian term.

\subsection{Quantum Lanczos}
The quantum Lanczos algorithm expands an eigenstate as a linear combination of states obtained via QITE:
\begin{equation}
\ket{\Psi} = \sum_{n=0}^{M}  c_n \ket{\psi(\beta_n)}
\label{eq:wfnql}
\end{equation}
Variational minimization leads to the generalized eigenvalue problem $\mathbf{Hc} = \mathbf{Sc} E$, where the matrix elements $(\mathbf{H})_{mn} = \bra{\psi(\beta_m) } \hat{\mathcal{H}} \ket{ \psi(\beta_n)}$  and $(\mathbf{S})_{mn} = \expval{\psi  (\beta_m) | \psi (\beta_n)}$, with $\beta_m = m \Delta \beta$, are obtained from the QITE subroutine.
A convenient feature of QL is that the matrix elements can be (approximately) evaluated in terms of the normalization coefficients $N$ and symmetric energy expectation values:
\begin{equation}
S_{mn} \approx \bra{\Phi}  e^{-m \Delta \beta \hat{A}} e^{-n \Delta \beta \hat{A}}  \ket{\Phi} = \frac{N(\beta_m)N(\beta_n)}{N^2(\beta_k)}
\end{equation}
and,
\begin{equation}
\begin{split}
H_{mn} \approx & \bra{\Phi}  e^{-m \Delta \beta \hat{A}}  \hat{\mathcal{H}} e^{-n \Delta \beta \hat{A}}  \ket{\Phi} \\ = & \frac{N(\beta_m)N(\beta_n)}{N^2(\beta_k)} \expval{ \hat{\mathcal{H}} }{ \psi(\beta_k) }
\end{split}
\end{equation}
where $2k = m+n$.
This is a significant simplification that allows the determination of all quantities needed for QLanczos \textit{without} using ancilla qubits.

\subsection{Quantum Krylov (QK)}
In quantum Krylov diagonalization, a general state is written as a linear combination of the basis $\{ \psi_n \}$ generated from real-time Hamiltonian evolutions at a given set of $s+1$ time steps $\{t_1, t_2, \ldots,t_s\}$ (typically chosen of the form $t_n = n \Delta t$) as
\begin{equation}
\ket{\Psi} = \sum_{n=0}^{s}  c_n \ket{\psi_n}= \sum_{n=0}^{s} c_n e^{-it_n \hat{\mathcal{H}}} \ket{\Phi}
\label{eq:wfnqk}
\end{equation}
Variational minimization of the energy of the state $\Psi$ leads to a generalized eigenvalue problem (like in the case of QLanczos) $\mathbf{Hc} = \mathbf{Sc} E$,
where the elements of the overlap matrix ($\mathbf{S}$) and Hamiltonian ($\mathbf{H}$) are given by $S_{mn} = \expval{\psi_m | \psi_n}$ and $H_{mn} = \bra{\psi_m} \hat{\mathcal{H}} \ket{\psi_n}$, respectively.
In \qft the quantum circuit used to approximate the time evolution is generated via Eq.~\eqref{eq:exp_pt2} using the Suzuki-Trotter decomposition
\begin{equation}
\label{eq:m_trotter_step2}
e^{-it\hat{\mathcal{H}}} = e^{-it \sum_\ell \theta_\ell \hat{P}_\ell} \approx \bigg( \prod_\ell e^{\frac{-it \theta_\ell \hat{P}_\ell}{r} } \bigg)^r
\end{equation}
with $r$ Trotter steps.
The quantum circuits used to evaluate $\mathbf{S}$ and $\mathbf{H}$ are implemented in \qft via a variant of the Hadamard test, requiring only an additional ancillary qubit\cite{aharonov2009polynomial}.
In \qft the time step $(\Delta t)$, number of time evolutions states $(s)$, and number of Trotter steps $(r)$ are all given as user-specified values.
We note that it is also possible to perform the exact time evolution operation (as opposed to those resulting from Trotterized dynamics) in \qft using sparse matrix operations.

\subsection{Multireference selected QK}

A selected multireference variant of QK (MRSQK) is also implemented in \textsc{QForte}.
The base procedure is identical to that of QK described in the above section, but several orthogonal reference states $\ket{\Phi_I}$ are included in the subspace and time evolved in order to improve numerical stability and target states with multireference character.
The MRSQK wave function is thus given by
\begin{equation}
\ket{\Psi} = \sum_\alpha  c_\alpha \ket{\psi_\alpha}= \sum_{I=1}^{d} \sum_{n=0}^{s} c_{I}^{(n)} e^{-it_n \hat{\mathcal{H}}} \ket{\Phi_I}
\label{eq:wfnmrsqk}
\end{equation}
In MRSQK, a preliminary single-reference QK calculation is performed in order to determine the set of important references.
The resulting single-reference QK wave function $\ket{\tilde{\Psi}} = \sum_n \tilde{c}_n \ket{\psi_n}$ is used to construct a list of determinants with importance value $P_\mu = |\langle\phi_\mu|\tilde{\Psi}\rangle|^2$, since the probability of measuring a determinant $\phi_\mu$ is equal to $P_\mu = |\langle\phi_\mu|\tilde{\Psi}\rangle|^2$.
In \qft, the quantity $P_\mu $ is approximated by measuring each element of the Krylov basis and estimating the total probability as a weighted sum over references via
\begin{equation}
P_\mu = |\sum_\alpha \langle \phi_\mu|\psi_\alpha \rangle c_\alpha|^2
\approx
\sum_\alpha |\langle \phi_\mu|\psi_\alpha \rangle |^2 |c_\alpha|^2
\label{eq:etaMA}
\end{equation}
Once formed, the list of the most important determinants is augmented to guarantee that all spin arrangements of open-shell determinants are included, and the $d$ most important references are used in the MRSQK subspace.
The number of references ($d$), the MRSQK time step $\Delta t_\mathrm{mr}$, the number of time evolutions per reference $(s)$, and the parameters for the preliminary single-reference QK calculation are all specified by the user at runtime.

\section{Representative timings}
\label{sec:timings}
By implementing an optimized low-level state-vector simulator, the algorithms implemented in \qft can be applied in a reasonable amount of time on small-sized molecular systems.
This enables the rapid development and generation of paper-quality benchmark data for new algorithms implemented in \qft.
To illustrate this point, we report timings for operations common to many quantum algorithms, such as application of circuits and evaluation of expectation values.
As an example, we consider the cost to generate a state-vector corresponding to a disentangled UCC ansatz (see Eq.~\ref{eq:ducc}) with particle-hole single and double excitations (dUCCSD) for a family of molecular hydrogen systems \ce{H2}--\ce{H10}.
The dUCCSD state is produced by applying the corresponding unitary $\hat{U}_\mathrm{SD}$ to a Hartree--Fock reference.
In this example, we employ a minimum basis set, corresponding to computations on 4--20 qubits.
We also consider the time required to evaluate the  Hamiltonian expectation value of the dUCCSD state via exact operator averaging: $\expval{\mathcal{H}_q} = \sum_\ell^{N_\mathrm{PS}} h_\ell  \expval{\hat{P}_\ell}$.    
Figure~\ref{fig:hn_timings} shows that \qft can accomplish such operations on the order of fractions of a second to a few minutes (on a laptop computer) for the systems considered here.
In particular, operations on systems with 6 electrons can be performed in the order of less than a second, enabling rapid testing of algorithms on relatively complex problems.

\begin{figure}[h!]
   \centering
   \includegraphics[width=3.0in]{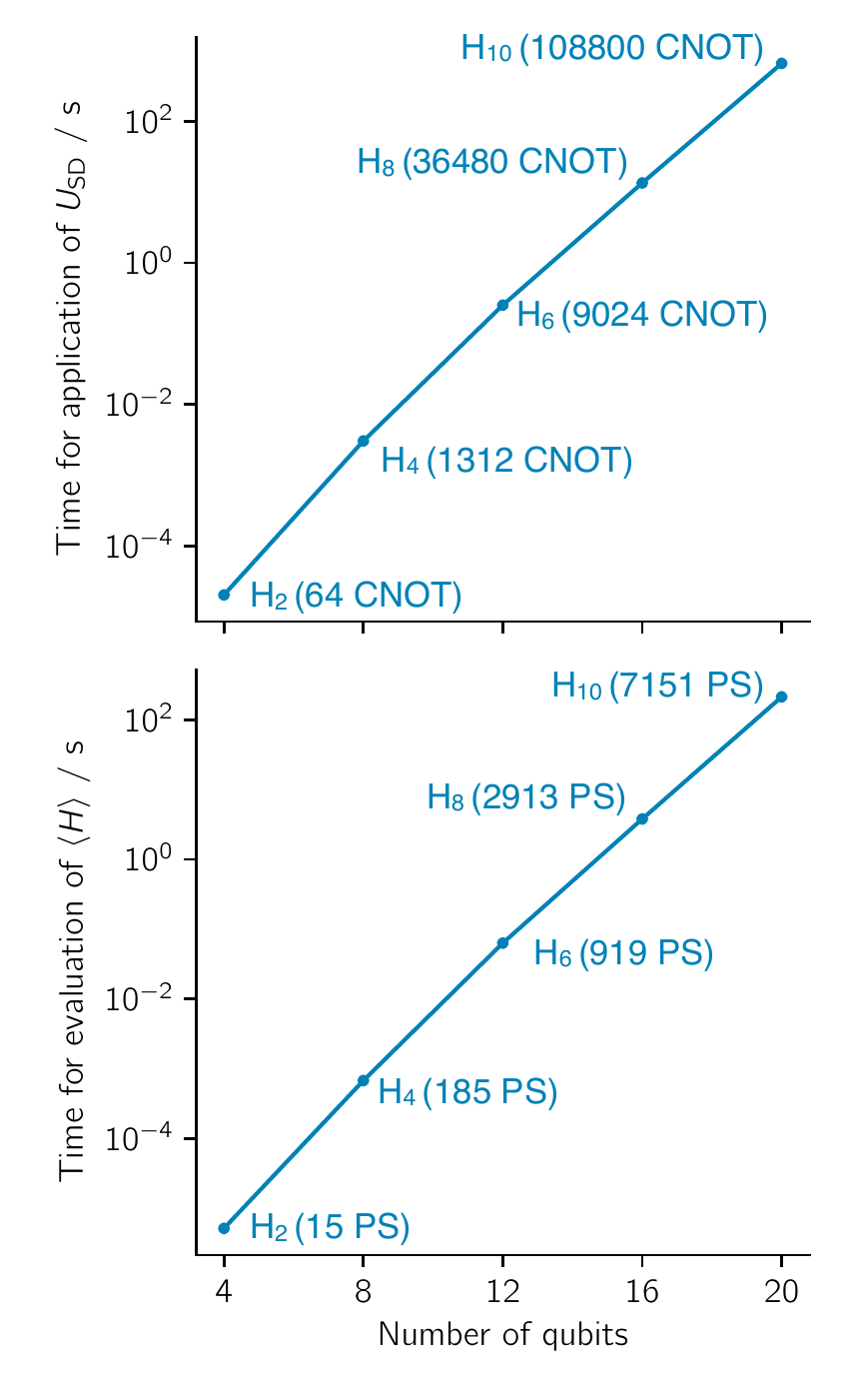}
   \caption{Timings for (top) the application of disentangled UCCSD circuits to a quantum states of increasing dimension, and for (bottom) the evaluation the molecular hydrogen Hamiltonians of increasing size (\ce{H2}--\ce{H10}). The number of $\rm{CNOT}$ gates in the dUCCSD circuits and the number of Pauli strings (PS) in the Hamiltonian are reported in parentheses beside the corresponding systems in the top and bottom plots, respectively. All cluster amplitudes in the dUCCSD circuits were initialized to 1.0. All timings shown were generated on a laptop computer with an Intel i5 3.1~GHz processor.}
   \label{fig:hn_timings}
\end{figure}

\section{Example: Developing new algorithms with \qft}
\label{sec:ex_usage}
In this section we discuss some examples of how \qft can be used to facilitate the implementation of new quantum algorithms.
We also then show how \qft can be used to produce comparative studies of different algorithms.
In addition to the example described below, \qft has several Jupyter-notebook tutorials available on topics ranging from basic API use, to detailed instructions for various algorithm implementations, to running jobs in a black-box fashion.

As an example, we consider the workflow necessary to experiment with a new VQE ansatz based on the theory of paired CC doubles.\cite{Limacher2013NewMean, Bulik2015CanSingle}
We will formulate the ansatz in the disentangled unitary coupled cluster form [Eq.~\eqref{eq:ducc}], such that operators entering into the resulting (disentangled) paired UCC doubles (p-dUCCD) ansatz maintain a seniority-zero trial state (include only contributions from closed-shell determinants).
We would also like to enforce that the ansatz include only particle-hole excitations/de-excitations, such that the final ansatz has the form:
\begin{equation}
\hat{U}_\text{p-dUCCD}(\bm{t}) = \prod_{i}^\mathrm{occ} \prod_{a}^\mathrm{vir} e^{ t_i^a (\cop{i_\beta} \cop{i_\alpha} \aop{a_\alpha} \aop{a_\beta} - \cop{a_\beta} \cop{a_\alpha} \aop{i_\alpha} \aop{i_\beta}) }
\end{equation} 
where the indices $i$, and $a$ pertain to occupied or virtual spatial orbitals, respectively, for a specified Hartree-Fock reference state.

The Python class structure in \qft can easily facilitate the implementation of the p-dUCCD-VQE variant.
To begin, we define a mixin class (pdUCCD) that defines only the function \code{ansatz\_circuit()}, responsible for returning the parameterized unitary circuit $\hat{U}_\text{p-dUCCD}(\bm{t})$ comprised of gates of the form shown in Fig.~\ref{fig:exp_pt} for the exponentials of Pauli-strings.
The \code{ansatz\_circuit()} function provided by the mixin class will be called by the function \code{AnsatzAlgorithm.energy\_feval()} which applies $\hat{U}_\text{p-dUCCD}(\bm{t})$ to a \code{Computer()} (initialized to the Hartree--Fock reference) and returns the energy expectation value (without noise by default).
Next, we define a new class \code{pdUCCDVQE}derived from the \code{VQE} abstract base class and the \code{pdUCCD} mixin class.
Once the \code{pdUCCDVQE} child class is defined, we simply need to define two functions: \code{run()} and \code{solve()}.
The \code{run()} function takes the user-defined parameters such as convergence thresholds, argument $D$ (number of layers), defines algorithm specific attributes such as the number of occupied and virtual orbitals, and calls all necessary subroutines to run the algorithm.

Finally, the \verb+solve()+ function calls the user-specified classical optimization algorithm (BFGS\cite{broyden1970convergence1, fletcher1970new, goldfarb1970family, shanno1970conditioning} by default), and defines: (i) the number of classical parameters used, (ii) the number of $\rm{CNOT}$ gates in the ansatz circuit (without considering advanced compilation techniques\cite{Hastings2015Improving}), and (iii) to total number of Pauli-string evaluations 
($\bra{\Phi_0} \hat{U}_\text{p-dUCCD}^\dagger(\bm{t}) \hat{P}_\ell \hat{U}_\text{p-dUCCD}(\bm{t})  \ket{\Phi_0}$).
An overview of the steps described above is given in Lst.~\ref{lst:levqe_class}.
We note that in practice \qft requires that derived classes define printing and attribute verification functions in addition to the three described above, but we do not show these in Lst.~\ref{lst:levqe_class} for brevity.

\begin{lstlisting}[language=Python, float=*, caption={Example Python code i) defining a mixin class \code{pdUCCD} responsible for the construction of the circuit for the p-dUCCD ansatz, and ii) defining a derived \code{pdUCCDVQE} class responsible for the optimization of the ansatz. Note that \qft requires that all derived classes additionally define printing and attribute verification functions in addition to those shown here.}, label={lst:levqe_class}]
from qforte import *
from scipy.optimize import minimize 

class pdUCCD:
    def ansatz_circuit(self, params):
        Kq = QubitOperator()

        for i in range(self._nocc):
            for a in range(self._nvir):
                mu = i*self._nocc + a
                k_mu = SQOperator()
                ia = 2*i
                ib = 2*i+1
                aa = 2*self._nocc + 2*a
                ab = 2*self._nocc + 2*a+1
                k_mu.add( 1.0*params[mu], [ab, aa], [ia, ib])
                k_mu.add(-1.0*params[mu], [ib, ia], [aa, ab])
                Kq.add(k_mu.jw_transform())

        U, phase = trotterize(Kq)
        return U

class pdUCCDVQE(pdUCCD, VQE):
    def run(self, optimizer='BFGS'):
        self._optimizer = optimizer
        self._nocc = int(sum(self._ref) / 2)
        self._nvir = int((self._nqb - sum(self._ref)) / 2)
        self._npar = int(self._nocc*self._nvir)
        self._params_0 = [0.0 for k in range(self._npar)]
        self.solve()

    def solve(self):
        res =  minimize(self.energy_feval, self._params_0, method=self._optimizer)
        self._final_res = res
        self._Egs = res.fun
        self._n_classical_params = len(res.x)
        self._n_cnot = self.build_Uvqc(res.x).get_num_cnots()
        self._Umaxdepth = self.build_Uvqc(res.x)
        self._n_pauli_trm_measures = self._Nl * res.nfev
\end{lstlisting}

After implementing the p-dUCCD-VQE method, we would like to compare it to other quantum algorithms, like VQE based on a conventional disentangled UCC ansatz.
Once an algorithm has been implemented, running jobs in \qft is very simple, as we only need to pass a molecule object to an algorithm constructor (see Sec.~\ref{sec:mol_class}).
For example, Lst.~\ref{lst:runing_h4} shows how to calculate the potential energy curve for \ce{H4} using the new p-dUCCD-VQE method, VQE with a disentangled UCC single and double excitations ansatz (dUCCSD-VQE), dUCCSD optimized via PQE (dUCCSD-PQE), and quantum Krylov diagonalization with a Hartree-Fock reference (QK).
\begin{lstlisting}[language=Python, float=*, caption={Example of how to compute a potential energy curve of the linear \ce{H4} molecule using both the \code{pdUCCDVQE} algorithm implemented in List.~\ref{lst:levqe_class} and various black-box algorithms implemented in \qft.}, label={lst:runing_h4}]
from qforte import *
import numpy as np

for r in np.linspace(0.5, 2.0, 30):

    # Specify geometry as a function of r.
    geom = [('H', (0., 0., 0.0)),
            ('H', (0., 0., 1*r)),
            ('H', (0., 0., 2*r)),
            ('H', (0., 0., 3*r))]

    # Run classical scf with Psi4 backend.
    mol = system_factory(build_type='psi4', mol_geometry=geom, basis='sto-3g')

    # Instantiate and run new VQE algorithm.
    pdUCCD_VQE = pdUCCDVQE(mol)
    pdUCCD_VQE.run()

    # Instantiate and run black-box algorithms.
    dUCCSD_VQE = UCCNVQE(mol)
    dUCCSD_VQE.run(pool_type='SD', opt_thresh = 1.0e-4)

    dUCCSD_PQE = UCCNVQE(mol)
    dUCCSD_PQE.run(pool_type='SD', res_vec_thresh = 1.0e-4)

    QK = SRQK(mol)
    QK.run(s=3)
\end{lstlisting}
After running the algorithms, it is possible to extract salient information such as the predicted energy,
and computational resource estimates.
Using the code in Lst.~\ref{lst:runing_h4}, we can produce the potential energy curve scans shown in Fig.~\ref{fig:levqe_ucc_comp} and obtain the computational resources estimates reported in Tab.~\ref{tab:levqe_ucc_comp}.
In this example, we conclude that although the p-dUCCD-VQE circuit is very compact (from a circuit depth and classical parameterization standpoint), it does not afford the same variational flexibility as the other dUCC-based methods or QK.
\begin{figure}[h!]
   \centering
   \includegraphics[width=3.0in]{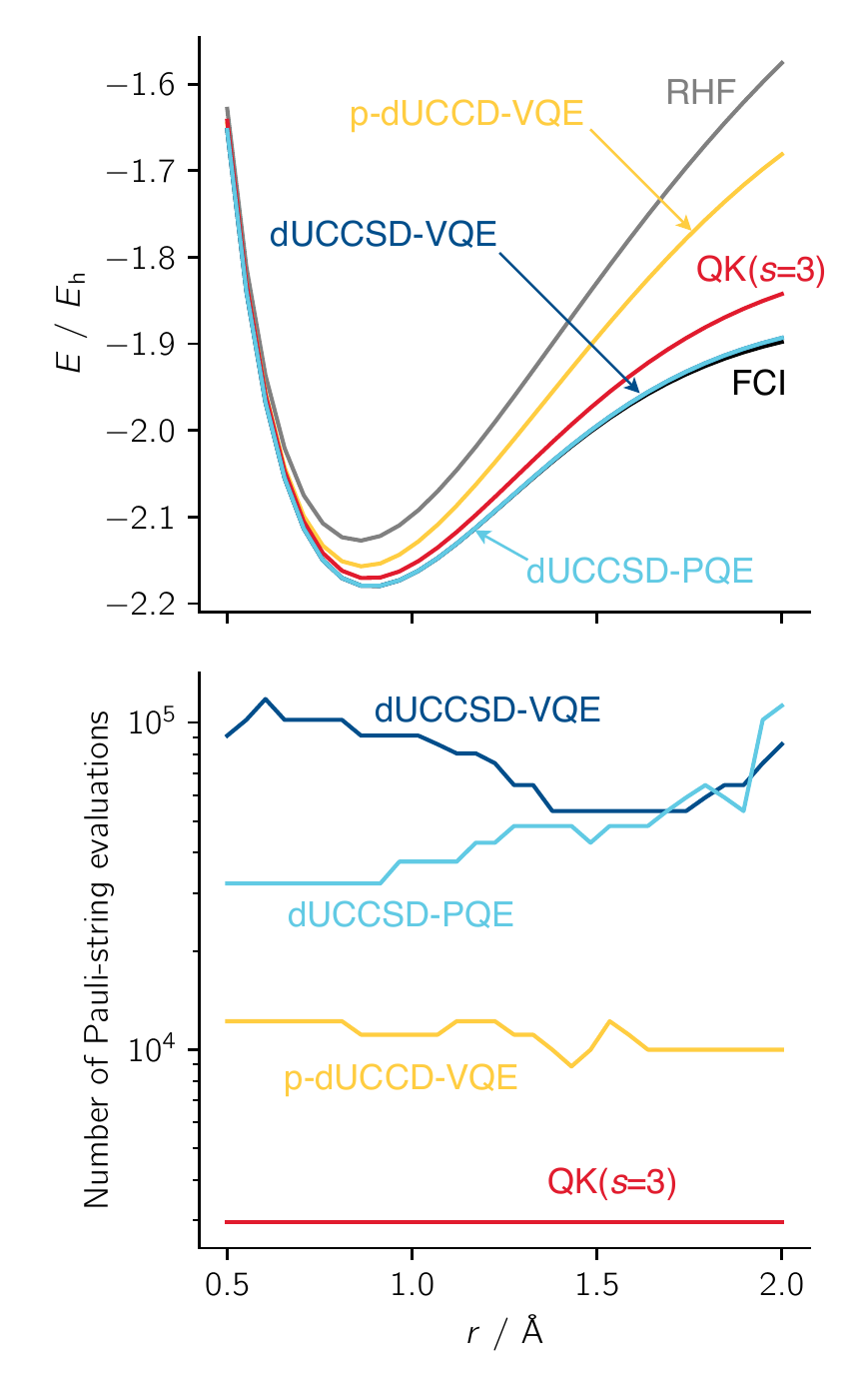}
   \caption{Ground state potential energy curve (top) and number of Pauli-string evaluations (bottom) computed with the p-dUCCD-VQE method implemented in Lst.~\ref{lst:levqe_class}, dUCCSD-VQE, dUCCSD-PQE, and QK with three time evolved basis states for the dissociation of linear \ce{H4} in a STO-3G basis. The QK calculation used a time step of $\Delta t = 0.5$~a.u. and a Trotterized time evolution circuit with a single Trotter step ($r=1$). The restricted Hartree-Fock (RHF) and FCI curves are also reported for reference.}
   \label{fig:levqe_ucc_comp}
\end{figure}

\begin{table}[!ht]
\centering
\renewcommand{\arraystretch}{1.2}
\caption{Computational resource estimates and mean signed error for p-dUCCD-VQE, dUCCSD-VQE, dUCCSD-PQE and QK(s=3, $\Delta t = 0.5$~a.u.) computed for the dissociation of linear \ce{H4} in a STO-3G basis.
$N_\mathrm{par}$ is the number of classical parameters used in the ansatz circuit for VQE/PQE or the dimension of the generalized eigenvalue problem for QK, $N_\mathrm{CNOT}$ is the number of $\mathrm{CNOT}$ gates in the ansatz (for VQE/PQE) or Trotterized time-evolution circuit (for QK), $\bar{N}_\mathrm{PSE}$ is the average number of Pauli-string evaluations over the entire potential energy curve, and MSE is the mean signed energy error (in m$E_\mathrm{h}$, with respect to the exact energy) computed over the entire potential curve (H-H near-neighbor bond distances in the range 0.5--2.0 \AA{}).}
\footnotesize
\begin{tabular*}{\columnwidth}{@{\extracolsep{\stretch{1.0}}}*{1}{l}*{4}{r}@{}}

    \hline

    \hline
     Method  & $N_\mathrm{par}$ & $N_\mathrm{CNOT}$ & $\bar{N}_\mathrm{PSE}$ & MSE (m$E_\mathrm{h}$)  \\
    \hline
    p-dUCCD-VQE         &  4	         & 192                                 &  11063 	& 80.922    \\
    dUCCSD-VQE         &  14	         & 736                               &  134804 	&    1.204    \\
    dUCCSD-PQE         &  14	         & 736                               &  94283 	&    1.204    \\
    QK                            &  4	         & 2656                             &  2972 	& 23.483    \\

    \hline
    
    \hline
\end{tabular*}
\label{tab:levqe_ucc_comp}
\end{table}

\section{Conclusion}
In this article we introduce the new open-source software package \qft. 
\qft aids the development and testing of quantum algorithms for molecular electronic structure, and has been used by our research group to implement the algorithms introduced in Refs.~\citenum{Stair_2020} and~\citenum{stair2021simulating}.
The ability of \qft to facilitate black-box calculations with a wide variety of quantum algorithms using only a classical electronic structure package as a dependency makes it an ideal tool for comparing quantum algorithms based on the accuracy of their outputs and their required computational resources.
Moreover, the easy-to-use basic components (implemented as C++ classes exposed in Python) combined with a simple Python class structure allow \qft to function as an excellent platform for implementing and testing new quantum algorithms.

In the future, we plan to expand the features and quantum algorithms implemented in \qft with the hope that its unique capabilities will provide a useful tool to other researchers.
For example, we plan to add support in the low-level layer of \qft for spin and point-group symmetries and enable the treatment of open-shell systems.
Another desirable feature is supporting fermionic encodings beyond the currently-available Jordan-Wigner transformation. 
On the front of the simulator performance, a significant speedup could be achieved by applying particle number and spin symmetry restrictions to the state vector, such that the application of fermionic operators can be performed efficiently using well-established quantum chemistry techniques,\cite{knowles1984new} extending the scope of routine computation to systems with more than 24 qubits.
Such an approach has recently been realized as a python implementation in the so-called fermionic quantum emulator,\cite{rubin2021fermionic} but has yet to be implemented in a compiled language.

On the algorithm side, we plan to extend the type of ans\"{a}tze supported in \qft (see, e.g., Refs.~\citenum{lee2018generalized,Ryabinkin:2018jw,wecker2015progress,foss2021holographic, haghshenas2021variational}).
We are also interested in implementing additional excited-state methods (see, e.g., Refs.~\citenum{mcclean2017hybrid,seki2021quantum,huggins2020non,ollitrault2020quantum}).
In addition to new algorithms and ans\"{a}tze, we also aim in the future to implement techniques for quantum resource reduction such as qubit tapering,\cite{bravyi2017tapering} Hamiltonian factorization,\cite{huggins2021efficient} and circuit compilation.\cite{Hastings2015Improving} 
It is our hope that \qft will become a valuable asset to the quantum simulation community.
We welcome feedback and contributions from any who wish to see their work represented in our package.

\begin{acknowledgement}
This work was supported by the U.S. Department of Energy under Award No. DE-SC0019374.
N.H.S. was supported by a fellowship from The Molecular Sciences Software Institute under NSF grant ACI-1547580.
\qft was predominantly developed as a Molecular Sciences Software Institute fellowship project.
N.H.S. would like to thank his mentor at the institute Jonathan Moussa for his thoughts and advice. 
In addition to development by N.H.S. and F.A.E., the authors acknowledge contributions to the \qft code from Nan He, Renke Huang, Jonathon Misiewicz, and Ilias Magoulas.
\end{acknowledgement}

\newpage
\bibliographystyle{achemso}
\bibliography{bibliography}
\end{document}